\documentclass[fleqn,10pt]{wlscirep}

\usepackage{dcolumn}
\usepackage{bm}
\usepackage{graphics} 

\usepackage[justification=justified]{caption}
\usepackage{circuitikz}
\usepackage{adjustbox}

\usepackage{url}

\title{Graphene Klein tunnel transistors for high speed analog RF applications}

\author[1,*,$\dagger$]{Yaohua Tan}
\author[1,$\dagger$]{Mirza M. Elahi}
\author[1]{Han-Yu Tsao}
\author[1,2]{K. M. Masum Habib}
\author[1]{N. Scott Barker}
\author[1,$\S$]{Avik W. Ghosh}
\affil[1]{Department of Electrical and Computer Engineering, University of Virginia, Charlottesville, Virginia 22904, USA}
\affil[2]{Intel Corp., Santa Clara CA 95054, USA.}
\affil[*]{\href{mailto:yt5x@virginia.edu}{yt5x@virginia.edu}}
\affil[$\S$]{\href{mailto:ag7rq@virginia.edu}{ag7rq@virginia.edu}}
\affil[$\dagger$]{these authors contributed equally to this work}

\begin{abstract}
We propose Graphene Klein tunnel transistors (GKTFET) as a way to enforce current saturation while maintaining large mobility for high speed radio frequency (RF) applications. The GKTFET consists of a sequence of angled graphene p-n junctions (GPNJs). Klein tunneling creates a collimation of electrons across each GPNJ, so that the lack of substantial overlap between transmission lobes across successive junctions creates a gate-tunable transport gap without significantly compromising the on-current. Electron scattering at the device edge tends to bleed parasitic states into the gap, but the resulting pseudogap is still sufficient to create a saturated output ($I_D-V_D$) characteristic and a high output resistance. The modulated density of states generates a higher transconductance ($g_m$) and unity current gain cut-off frequency ($f_T$) than GFETs. More significantly the high output resistance makes the unity power gain cut-off frequency ($f_{max}$) of GKTFETs considerably larger than GFETs, making analog GKTFET potentially useful for RF electronics. Our estimation shows the  $f_T$/$f_{max}$ of a GKTFET with 1 $\mu$m channel reaches 33 GHz/17 GHz, and scale up to 350 GHz/53 GHz for 100 nm channel (assuming a single, scalable trapezoidal gate). The $f_{max}$ of a GKTFET is 10 times higher than a GFET with the same channel length.
\end{abstract}
\begin{document}

\flushbottom
\maketitle
%
%
\thispagestyle{empty}
\section{\label{sec:introduction}Introduction}

Graphene-based devices have long promised exciting applications, from interconnects and transparent electrodes to gas sensing. However, their gaplessness compromises our ability to gate 
these devices as an efficient electronic switch. 
For instance, graphene is a promising channel material for radio frequency (RF) applications \cite{schwierz2010graphene,schwierz2013graphene,fiori2014electronics,chauhan2011assessment} due to its intrinsic high carrier mobility and long mean free path \cite{bolotin2008ultrahigh, chen2008intrinsic, dean2010boron, morozov2008giant}. In fact, graphene RF devices have been reported to achieve $f_T$'s larger than 300 GHz for sub-100 nm channels \cite{wu2012state,cheng2012high}. 
However, the gaplessness of graphene makes its output resistance low, arising from the lack of any current saturation.  Consequently, the power gain cut-off frequency $f_{max}$ of most of the reported GFETs are much lower than their $f_T$,and does not scale with channel length\cite{schwierz2010graphene, schwierz2013graphene, fiori2014electronics} {(Fig \ref{fig:summary_of_fT_fmax} (f))} because of the non-scalability of the dominant contact resistances.  

Efforts to improve the $f_{max}$ of GFETs have focused on reducing the input resistance and introducing current saturation. Recent work by Guo {\it{et al.}} \cite{guo2013record} showed an improved $f_{max}$ in GFETs  by significantly reducing the gate resistance  using a T-shaped gate. To obtain current saturation in GFETs, an energy bandgap can in principle be introduced in graphene, such as by  applying symmetry breaking strain \cite{ni2008uniaxial} or using quantum confinement in graphene nanoribbons and nanotubes  \cite{han2007energy}. Furthermore,  scattering process in a long graphene channel can also introduce natural current saturation \cite{perebeinos2010inelastic}.
However, those band gap opening mechanisms significantly reduce the carrier mobility due to the distorted bandstructure or  carrier scattering events \cite{meric2008rf}.  
Thus a technique which introduces a transport gap in graphene without degrading the carrier mobility \cite{sajjad2011high, sajjad2013manipulating} would be quite unique and highly desirable for graphene-based RF applications.

Over several publications in the past, we have proposed an alternative way of introducing gaps into graphene, through the employment of p-n junctions as collimator-filter pairs. A GPNJ is an angle dependent momentum filter driven by the Klein tunneling of Dirac electrons \cite{beenakker2008colloquium,allain2011klein,katsnelson2006chiral, low2009conductance, sajjad2012manifestation}. As a result, the GPNJ creates gate-tunable transmission gaps instead of energy band gaps, so that states are available for conduction in the ON state but removed for the OFF state. 
The underlying physics of Klein tunneling in tilted GPNJ has been demonstrated not only by experiments\cite{sutar2012angle}, but also by theoretical calculations using quasi-analytical model\cite{sajjad2013manipulating}, numerical models using semiclassical ray tracing\cite{elahi2016current}, 
and fully quantum NEGF\cite{sajjad2013manipulating}. 
Multiple papers in the past have suggested using the GKTFET as a digital switch \cite{sajjad2011high, sajjad2013manipulating, jang2013graphene, wilmart2014klein}. Initial calculations \cite{sajjad2011high} treating the angled junctions as independent transmitters estimated ultrahigh ON-OFF ratios in excess of 10$^4$, while the gate tunability of the transport gap predicted a subthreshold swing that beats the fundamental Boltzmann limit. However, in practice the ON-OFF ratio of these devices is seen to be compromised by recurrent momentum scattering of rejected electrons at the device edges, which typically redirects leakage states into the transmission gap and limits the experimentally measured ON
-OFF ratio to anywhere between 1.3 \cite{morikawa2017dirac} to 12. In other words, the predicted gap readily turns into a pseudogap because of parasitic scattering events.

In this work, we propose to use the pseudogap in GKTFETs for RF applications to overcome the lack of current saturation in traditional GFETs. To understand the characteristics of GKTFETs, we performed semiclassical ray tracing calculations coupled with analytical models for Klein tunneling to model electron transport in GKTFETs \cite{chen2016electron}. The critical device parameters for a given geometry are extracted from finite element electrostatic calculations in order to estimate the cut off frequencies.
According to our calculations, we argue that even a pseudogap suffices to allow GKTFETs to have distinct current saturation \cite{elahi2016current} and considerably larger output resistance $r_0$ than conventional GFETs, in fact, in excess of their contact resistances. In the process, the mobility in a GKTFET is not significantly degraded because the transmission gap dominates only for the OFF state and is kept just small enough in the ON state to still allow saturation. We expect the GKTFET can reach a $f_T$ of 33 GHz in a 1 $\mu$m channel device, and scale up to  350 GHz at 100 nm channel length assuming ideal single gate scaling. The $f_{max}$ of GKTFET can reach 17 GHz in a device with a 1 $\mu$m channel  and 53 GHz at 100 nm length, which is more than 10 times higher than that of GFETs at a comparable channel length. Higher $f_{max}$ of 49 GHz(1 $\mu$m) and 158 GHz (100 nm) can be reached if the gate resistance of GKTFET can be significantly reduced by reducing the gate input resistance, such as with a T-Gate.

\begin{figure}[h]
\centering
	\includegraphics[width=0.9\linewidth]{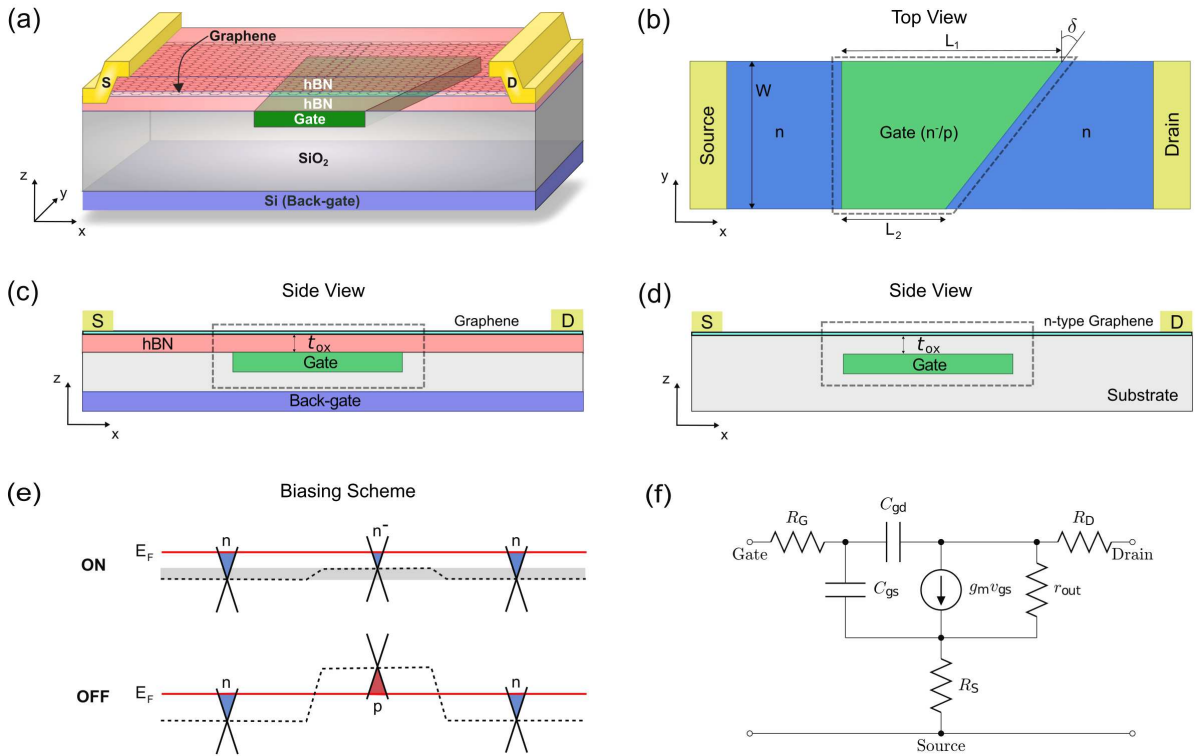}
	\caption{\textbf{Klein tunnel FET based on dual tilted graphene p-n junctions (GPNJ)}. (\textbf{a}) 3D Schematic. First SiO$_2$ is grown on top of Si back-gate, then polysilicon/graphite wedge shaped gate (local) is deposited/stamped. A graphene flake is sandwiched between hBN and then transferred on top of SiO$_2$ to make sure high quality graphene is achieved \cite{dean2010boron}. (\textbf{b}) Top view. Gate (local) controls the charge concentration in the central green region. In the OFF state (n-p-n), two back-to-back GPNJs are formed. The left GPNJ acts as a collimator and right GPNJ acts as a filter.  In the ON state (n-n$^-$-n), GPNJ on the right is tilted by angle $\delta=45^\circ$ with respect to the left one. {Here we approximated the potential profile changes linearly across the junction \cite{supplementary}.} In this work, the device has an average gate length of $(L_1+L_2)/2$ = 1 $\mu$m ($L_1$ = 1.5 $\mu$m, $L_2$ = 0.5 $\mu$m) and a width of $W$ = 1 $\mu$m. Gate dielectric is equivalent to 5 nm SiO$_2$ (EOT = 5 nm). (\textbf{c}) Side view for electrostatic doping by gate. (\textbf{d}) Side view for chemical doping case where back-gate is not needed for controlling regions other than ones covered by local gate. The essential part of the device is shown in dashed boxe, where the Klein tunneling effect near GPNJs dominates in (b, c, d). (\textbf{e}) Gate biasing scheme for ON and OFF state. Gray region corresponds to the energy range of the transmission gap in the ON state. (\textbf{f}) Equivalent small signal circuit.}
	\label{fig:device}
\end{figure}

\section{\label{sec:device}Methods}
\begin{figure}[h]
	\centering
	\includegraphics[width=0.95\textwidth]{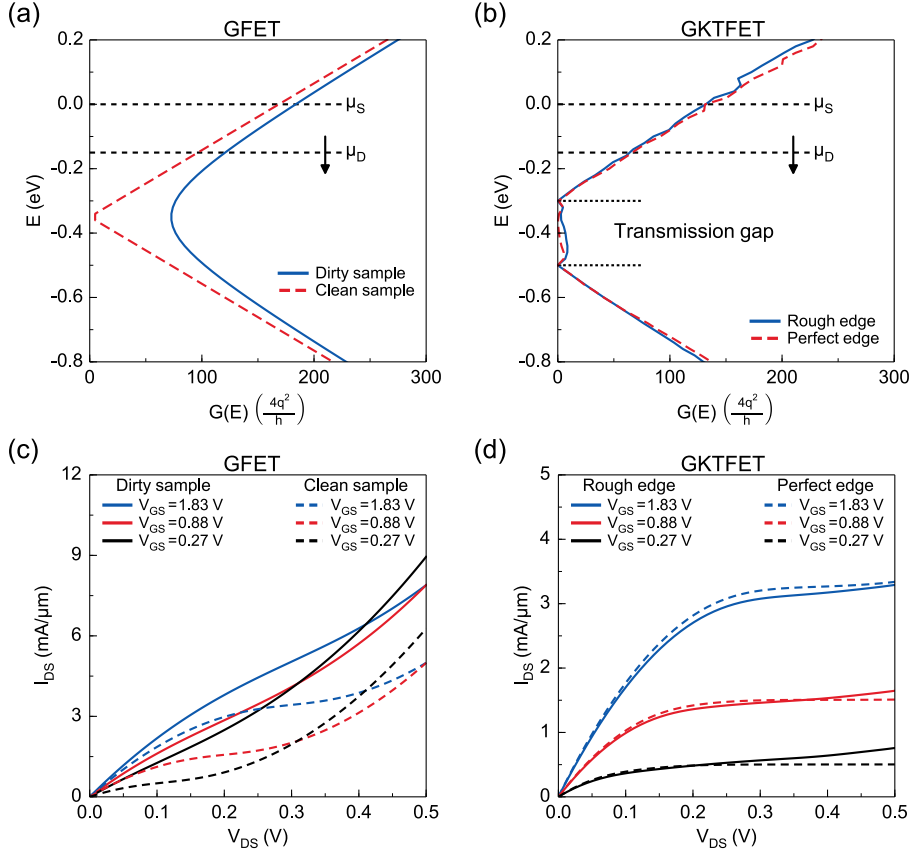}
	\caption{\textbf{Conductance and output characteristics}. (\textbf{a}) Energy resolved conductance for GFET (on-state). $G(E) \propto |E|$ corresponds to the Dirac cone-like band structure of clean sample (dash), and $G(E) \propto \sqrt{E^2+2\sigma^2/\pi}$ in dirty sample (solid)\cite{sajjad2015quantum}. (\textbf{b}) Energy resolved conductance for GKTFET (on-state). A clear transmission gap can be observed unlike GFET in (a). $G(E)$ in the transmission gap is slightly non-zero due to the edge reflections even with the perfect edge. Adding edge roughness creates more states inside the transmission gap. (\textbf{c}) $I_{DS}$ vs. $V_{DS}$ for GFET. $I_{DS}$ in dirty graphene sample (solid) is a linear function of $V_{DS}$ whereas clean graphene sample (dash) shows one point saturation. (\textbf{d}) $I_{DS}$ vs. $V_{DS}$ for GKTFET. GKTFET shows obvious current saturation in both cases with (solid) and without (dash) edge roughness. With edge roughness, it shows slightly larger slope in the saturation region which in turn reduces $r_{out}$ due to increment of states inside the transmission gap. Gate voltages are calculated considering quantum capacitance where 0.3 V, 0.2 V, and 0.1 V are dropped in channel respectively for gate voltages mentioned both in (c) and (d).
	}
	\label{fig:transmission}
\end{figure}

Klein tunneling across a graphene p-n junction is driven by the conservation of pseudospin, which in turn is set by the phase coherent superposition of the dimer p$_z$ basis sets. The GPNJ acts as
an efficient momentum filter, transmitting electrons that are injecting perpendicularly to the junction regardless of the barrier height across the junction. A GPNJ with a graded junction potential further filters the non-normal electrons aggressively, as those electrons see an angle dependent tunnel barrier akin to the cut-off modes in a rectangular waveguide \cite{cheianov2006selective, chen2016electron}. The device considered in this work contains two back-to-back GPNJs  controlled by {two gates including a trapezoidal gate and a back gate}, as shown in Fig.\ref{fig:device}. (a). The GPNJ on the right has a tilted angle $\delta (= \pi/4)$ with respect to the left one \cite{sajjad2013manipulating}.
In the OFF state,  n-p-n  regions are formed in the channel by applying proper gate biasing. These two back-to-back p-n junctions will collectively turn off the current. The left GPNJ serves as a collimator which blocks most of the incoming electrons except for those incident perpendicular to it; the second tilted GPNJ further blocks the electrons coming from the collimator allowing in turn only  electrons perpendicular to itself to pass. OFF state is thereby achieved through sequential momentum filtering when the angle of the second junction exceeds the critical angle at the first junction. The average gate length considered in this work is 1 $\mu$m with a split length of $d_{1,2}$=80 nm {}. 
For this paper, we assume a gate oxide with an equivalent SiO$_2$ thickness of 5 nm (EOT=5 nm).  
{In the ON state, the three regions in the graphene channel are held as n-n$^-$-n, so there is no angular filtering of electrons in the Fermi window between $\mu_S$ and $\mu_D$ for low $V_{DS}$. Filtering exists for portion of energy window (transmission gap shown in Fig. \ref{fig:transmission}b) which comes into act for high $V_{DS}$}. The small transmission gap exists due to slight differential doping (n-n$^-$). This leads to current saturation in the ON state. In the OFF state, we move the polarity of the central gate to n-p-n where the gap increases substantially and the current drops. 

It should be noted that the GKTFET proposed here is designed to establish proof-of-concept. In practice the geometry needs to be optimized keeping in mind the fabrication techniques, 
considering different approaches such as electrical gating (Fig.\ref{fig:device} (b)) or contact-induced doping\cite{cayssol2009contact}/chemical doping (Fig.\ref{fig:device} (c)) \cite{brenner2010single,tang2010raman} to create the side gated regions (blue n-doped regions in Fig.\ref{fig:device} (a)). According to our finite element electrostatic calculations using Ansoft Maxwell, the side gate in (Fig.\ref{fig:device} (b)) at the drain end (back gate) introduces a large parasitic capacitance. This extra capacitance will possibly compromise the cut-off frequencies if it is AC connected to the ground directly. Extra care should be taken to get rid of the effect of this capacitance, as we discuss later in this paper. Compared with electrostatic side gate doping, the chemical doping shown in Fig.\ref{fig:device} (c) does not suffer from these large gate capacitances. However chemically doped graphene has lower carrier mobility. In Fig.\ref{fig:device}, we showed a proposed device with buried gates 
that was the basis for the calculations in this paper
; however, it is worth looking at alternate geometries, such as top gates \cite{guo2013record,liao2010high}, with the associated design trade-offs.

In our proposed device, the total current $I_{DS}$ across the GKTFET can be estimated by the Landauer equation.
\begin{equation}
I_{DS} = \frac{2q}{h} \int { T_{av}M \Bigl[f(\mu_S) - f(\mu_D) \Bigr] dE}
\end{equation}
where $M$ is the number of modes, $T_{av}$ is their mode-averaged transmission, $q$ is the charge of the electron,  $h$ is the Planck's constant, $f$ is the Fermi-Dirac distribution, and $\mu_{S,D}$ are the bias-separated electrochemical potentials in the source and drain.
The mode-averaged transmission $T_{av}$ and number of modes $M$ at energy $E$ are controlled by the potential drops on the channel $V_{GS}$. The resulting transconductance $g_m$ can be written as
\begin{eqnarray}\label{eq:gm_analytical}
g_m   = \frac{\partial I_{DS}}{\partial V_{GS}}& \propto &  \int { \frac{ \partial \left( T_{av}M \right) }{\partial V_{GS}}\Bigl[f(\mu_S) - f(\mu_D) \Bigr] dE} \\
& \approx & T_{av}M |^{\mu_D}_{\mu_S}
\end{eqnarray} 
In the ON state, the GKTFET has a small transmission gap around the Dirac point 
so that its mobility and $g_m$ are expected to resemble a pristine GFET with the same dimensions.
The presence of a transmission gap will, however, cause the current to saturate when the drain electrochemical potential $\mu_D$ moves towards the Dirac point and enters the transmission gap. In contrast, the $g_m$ of ultra-clean GFETs has just a single point saturation precisely when $\mu_D$ hits the Dirac point since there is no gap in pristine graphene. This feature can be seen later in Fig.\ref{fig:gm_rout}.
The output resistance $r_{out}$ can be estimated by
\begin{equation}\label{eq:output_resistance}
r_0 = \frac{\partial V_{DS}}{\partial I_{DS}} \propto  \left( \int MT_{av} \frac{\partial f(\mu_D)}{\partial \mu } dE \right)^{-1}.
\end{equation} 
{From eq (\ref{eq:output_resistance} ), it can be seen that the output resistance depends on the modes inside the band gap 
in quasi-ballistic limit.}
A perfect energy gap in principle leads to infinite output resistance because $MT_{av}=0$ in the gap, while any states inside the gap due to imperfections (such as scattering, defects) will lead to a  finite output resistance. In our proposed devices, the output resistance is limited primarily by the edge reflection and carrier scattering.

In this work, we simulate the GKTFET using a semiclassical ray-tracing method coupled with analytical equations for chiral tunneling across the junctions \cite{chen2016electron}. Standard quantum transport methods like the non-equilibrium Green's function formalism (NEGF)
\cite{sajjad2015quantum,sajjad2013atomistic}, are computationally quite expensive for GKTFETs with sizes between few hundreds of nanometers to micro-meters, and moreover, bring in spurious interference effects that are irrelevant at room temperature. In contrast, the ray-tracing method \cite{beenakker1989billiard, milovanovic2013spectroscopy, milovanovic2014magnetic} coupled with a well-benchmarked quantum tunneling model across junctions can be applied readily to devices at high voltage bias with large areas and complicated geometries including a sequence of multiple reflections, and has shown excellent agreement with recent experiments on GPNJs \cite{chen2016electron}. In our approach, electrons are thrown from the source randomly with injection angles following a cosine distribution (the angular distribution of the quantized transverse wave-vectors). The average transmission ($T_{av}$) across the junctions is then calculated by counting electrons that successfully reach the drain. The analytical transmission probability for each electron across the junction is a simple generalization of Gaussian filtering $T \sim \exp^{-\pi k_F (d/2) \sin^2 \theta }$ established in \cite{cheianov2006selective} and   extended now to an asymmetrically doped junction \cite{sajjad2013manipulating}.
Here, $d$ is the split length of the junction,  $\theta$ is the incident angle of the electron, and  $k_F$ is the Fermi vector on both sides.

In our calculations, we considered the cases of GPNJ with perfect edges as well as rough edges.  Fig. \ref{fig:transmission} shows the integrated transmission of (a) bulk graphene  and (b) GKTFET, both for clean vs. dirty sample (charge puddles for bulk and edge roughness for GKTs).  
We see that pristine graphene has no band gap and its density of states $D(E) \propto |E|$ for a clean sample,  while $D(E) \propto \sqrt{E^2+2\sigma^2/\pi}$ for a dirty sample \cite{sajjad2015quantum}, with $\sigma^2 \approx 2\hbar^2v_F^2n_{imp}+C$ describing the contribution of charge puddles in washing out the Dirac point through spatial averaging. We use a typical impurity density in a dirty sample with $n_{imp}=1\times$10$^{12}$  cm$^{-2}$ in this work \cite{sajjad2015quantum}. In contrast with GFETs,  GKTFETs have a distinct transmission gap, as indicated in Fig. \ref{fig:transmission} (b). Indeed, very few modes appear in the transmission gap. These gap states arise from edge reflection of electrons rejected by the second junction, a process that redirects them towards the drain. Ultimately these states contribute to the  leakage current in the off-state and lead to a finite $r_{out}$. Our calculation indicates that perfect edges in the GKTFET can reduce the leakage current by 20 to 40 times compared with GKTFET with rough edges in a 1 $\mu$m wide device. In our calculation, the edge roughness introduces a random reflection angle with a variance of $\sigma = 18^\circ$.  

\section{\label{sec:RF Parameters}Results}
Fig. \ref{fig:transmission} (c) and (d) shows the $I_{DS}$-$V_{DS}$ characteristics of GFETs and GKTFETs. In each case, the dashed lines are for clean samples while the solid lines include imperfections. It can be seen clearly that an ultraclean GFET  shows $I_{DS}-V_{DS}$ with single-point saturation, while a GFET with a dirty sample shows a quasi-linear $I_{DS}-V_{DS}$ due to spatial averaging that washes out the Dirac point. In contrast, GKTFETs with both perfect and rough edges show a clear current saturation due to the presence of a pseudogap. The rough edges in GKTFETs lead to only a marginally smaller $r_{out}$ because of the increase of $MT_{av}$ in the transmission gap, which can be understood using eq.(\ref{eq:output_resistance}). 

\begin{figure}[h]
	\centering 
	\includegraphics[width=0.95\textwidth]{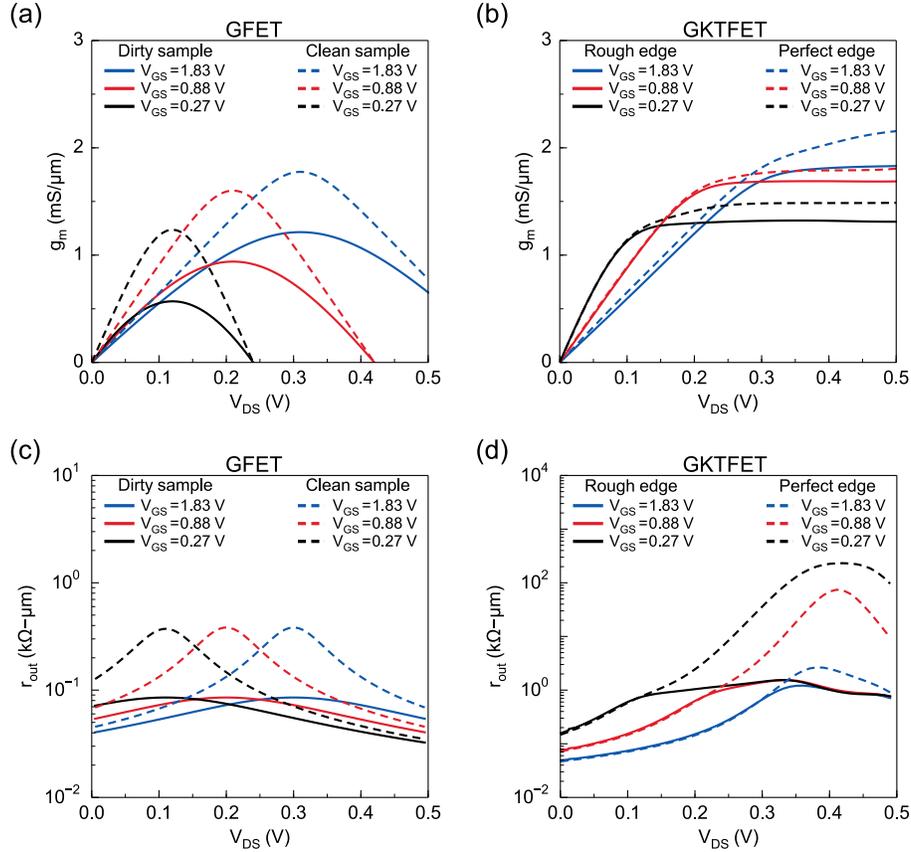}
	\caption{\textbf{Transconductance ($g_m$) and output resistance ($r_{out}$) for GFET and GKTFET}. (\textbf{a}) $g_m$ vs. $V_{DS}$ in GFET for dirty (solid) and clean (dash) sample where $g_m$ is between 0.5 to 1.5 mS/$\mu$m at saturation points. (\textbf{b}) $g_m$ vs. $V_{DS}$ in GKTFET with (solid) and without (dash) edge roughness where effect of edge roughness is not significant on $g_m$. (\textbf{c}) $r_{out}$ vs. $V_{DS}$ in GFET for dirty (solid) and clean (dash) sample where $r_{out}$ is found to be around 0.1 k$\Omega$-$\mu$m for dirty sample. (\textbf{d}) $r_{out}$ vs. $V_{DS}$ in GKTFET with (solid) and without (dash) edge roughness. Although $r_{out}$ in GKTFET reduces from 10-100 k$\Omega$-$\mu$m to about 1 k$\Omega$-$\mu$m due to edge roughness, still in both cases, output resistances ($r_{out}$) are greater than the ones for GFET in (c) utilizing transmission gap.}
	\label{fig:gm_rout}
\end{figure}
Fig . \ref{fig:gm_rout} show the $g_m$ and $r_{out}$ of GFET and GKTFETs. The $g_m$ of GFETs reach 0.5 to 1.5 mS/$\mu$m (each gated region is 1 $\mu$m long in our simulation with linear transition length of 80 nm (split length, d) each), while the GKTFET turns out to have a slightly higher $g_m$ of 1 to 2 mS/$\mu$m. The output characteristic however proves more dramatic than the transfer characteristic. The GFET shows a very low $r_{out}$ of 0.1 k$\Omega$-$\mu$m for dirty samples, and only around $r_{out} \sim$ 0.3 k$\Omega$-$\mu$m for clean samples at saturation $V_{DS}$ for all gate biases, dropping rapidly for other $V_{DS}$ values. At zero temperature, $\frac{\partial f(\mu_2)}{\partial \mu_2 } = \delta(E-\mu_2)$ in eq (\ref{eq:output_resistance}), $r_{out}=\infty$ as $MT_{av}=0$ at the Dirac point. At finite temperature $r_{out}$ drops to a finite value because $\frac{\partial f(\mu_2)}{\partial \mu_2 }$ has a non-vanishing spread of $kT$. Compared with GFETs, the GKTFET shows much higher $r_{out}$ of 1 k$\Omega$-$\mu$m even with edge roughness. GKTFET with perfect edges shows even higher $r_{out}$ values that can reach 50 to 100 k$\Omega$-$\mu$m. Furthermore, the saturation region corresponds to a $V_{DS}$ in the range of 0.1 to 0.3 V instead of one point saturation.

\begin{figure}[h]
	\centering
	\includegraphics[width=0.95\textwidth]{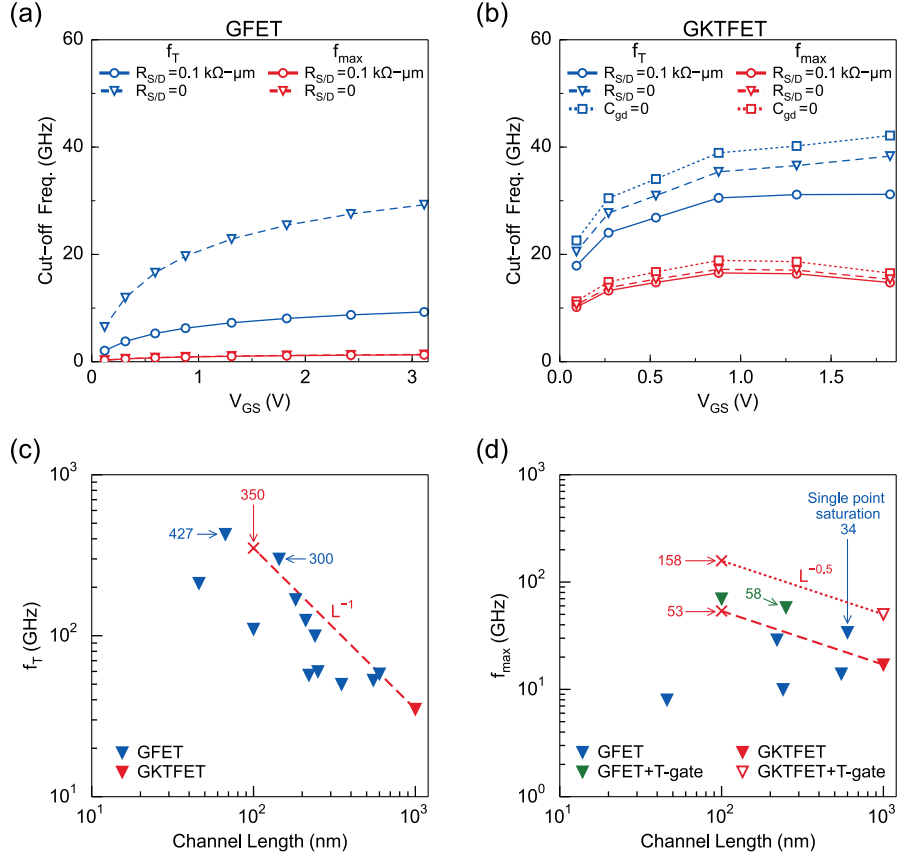}
	\caption{\textbf{Cut-off frequencies for GFET and GKTFET}. (\textbf{a}) $f_T$ and$f_{max}$ for GFET. The $f_{max}$ of GFET is significantly smaller than $f_T$ because of the small $r_{out}$. (\textbf{b}) $f_T$ and$f_{max}$ for GKTFET. $f_T$ and $f_{max}$ of GKTFET reach their maximum at the saturation region which range over 0.1 V to 0.3 V. Due to high output resistance,the max $f_{max}$ in GKTFETs is about 50\% of the maximum $f_T$. (\textbf{c}) $f_T$ vs. channel length of GKTFET compared to reported GFETs. (\textbf{d}) $f_{max}$ vs. channel length of GKTFET compared to reported GFETs.	
GFETs' data are from Ref. \cite{cheng2012high,guo2013record,lin2009development,lin2010100,meric2011high,liao2010high}. Reported $f_T$s of GFETs are roughly inversely proportional to channel length, while the $f_{max}$s do not show this trend due to low output resistance. For ideally scaled GKTFET with 100 nm channel length, $f_T$ is expected to reach 350 GHz (shown by red cross in (c)), which is comparable to the highest reported $f_T$s in GFETs (mentioned in the figure). The scaling of $f_{max}$ of GKTFET follows $L^{-0.5}$ ideally. For a GKTFET with 100 nm channel, the $f_{max}$ is expect to reach 53 GHz, and even high $f_{max}$=158 GHz can be expected if T-gate technique is used to reduce gate resistance (both shown by red cross in (d)).}
	\label{fig:summary_of_fT_fmax}
\end{figure}
To estimate the RF properties such as $f_T$ and $f_{max}$ of graphene-based RF devices, we consider the equivalent circuit for AC signals, shown in Fig.\ref{fig:device} (e).
 This structure assumes that the dominant capacitance is from the central gate that is swung between p and n polarities, while the side gated regions have lower capacitance (we discuss this point later in the paper) 
For this equivalent circuit, the $f_T$ and $f_{max}$ can be estimated by \cite{rutherglen2009nanotube}.
\begin{equation}\label{eq:f_T}
f_{T} = \frac{g_m}{2\pi\left(C_{gs}+C_{gd}\right)}
\frac{1}{ \sqrt{ \left[ 1+g_0(R_S+R_D)\right]^2 - (g_0R_S)^2} }
\end{equation}with $g_0 = g_{ds}+g_m\left[C_{gd}/(C_{gs}+C_{gd})\right]$
and
\begin{equation}\label{eq:f_max}
f_{max} = \frac{f_T}{2\sqrt{ \frac{R_G+R_S}{r_{ds}} + 2\pi f_T R_G C_{gd}}}.
\end{equation}
To illustrate the impact on $f_T$ and $f_{max}$ from improved $g_m$ and $r_{out}$ of GKTFET in comparison with GFETs, we use $C_{gs}$=6.9 pF/mm, $C_{gd}$=0.7 pF/mm and $R_{G}$=1 k$\Omega$-$\mu$m for both transistors. However, it is worth re-emphasizing that the parameters are strongly dependent on the device geometry, for instance, 
the $C_{gd}$ of the GKTFET in Fig.\ref{fig:device} (b) is in fact negligible by finite element electrostatic calculation using Ansoft Maxwell. We accordingly choose an experimentally achievable ratio of $C_{gd}/C_{gs}$=0.1 \cite{guo2013record} in the following calculations of $f_T$ and $f_{max}$. {It should be noted that our calculations for $f_T$ and $f_{max}$ using small $C_{gd}$ are only valid for saturation region.}

{Fig. \ref{fig:summary_of_fT_fmax} show the peak value of $f_T$ and $f_{max}$ of the GFET and GKTFET.} The $f_T$  reaches 9.3 to 29.3 GHz in GFETs with a channel length of $L_{channel}$=1 $\mu$m (better contact resistances and smaller $C_{gd}$ gives higher $f_T$). It is known that the $f_T$ in pristine GFETs is inversely proportional to $L_{channel}$ \cite{rutherglen2009nanotube}. Projecting accordingly, a channel length of 100 nm leads to an expected max $f_T$=100 to 300 GHz for a conventional GFET, which agrees with the published literature\cite{cheng2012high,guo2013record,lin2009development,lin2010100,meric2011high,liao2010high}. 
The $f_{max}$ of the GFET reaches only 1.3 GHz, i.e., 14\% of $f_{T}$ because of its small output resistance.  
Compared with the ideal case where $R_{S/D}$= 0, the $f_{max}$ of GFET with larger $R_{S/D}$=0.1 k$\Omega$-$\mu$m is reduced by $5\%$, while the peak $f_T$ is reduced by 69\%.  

Compared with GFETs, the $f_{T}$  of the GKTFET is larger due to a larger $g_m$ arising from the opening of the transport gap and the resulting variation in density of states over the finite temperature window. More noticeably, the $f_{max}$ and $f_{max}/f_T$ ratio in GKTFET are significantly higher due to the current saturation arising from the engineered pseudogap. Fig . \ref{fig:summary_of_fT_fmax} (b) and (e) show that the GKTFET with 1 $\mu$m channel length reaches a $f_T$ of 31 GHz and $f_{max}$ of 17 GHz. 
The $f_{max}$ is 13 times larger than that of GFET. Furthermore, the contact resistance has a much weaker impact on $f_T$ in GKTFET - in fact, 0.1 k$\Omega$-$\mu$m $R_S$ and $R_D$ reduces the $f_T$ by only $\sim$ 10-20\%. The impact of $R_S$ and $R_D$ to $f_T$ in both GFETs and GKTFET is determined by the factor $g_{ds}\left(R_S+R_D\right)$, $g_{ds}=1/r_{out}$, as we see in the denominator of equation (\ref{eq:f_T}). The large output resistance $r_{out}$ of GKTFET weakens the  influence of $R_S$ and $R_D$ on the $f_T$.  In Fig.\ref{fig:summary_of_fT_fmax} , we also show the $f_T$ and $f_{max}$ in the limit of $C_{gd}$=0 by the dashed lines. It can be seen that the small $C_{gd}$ leads to a 30\% increase of the max $f_T$ and 10\% increase of max $f_{max}$.

\section{Discussion}
 We have shown that a 1 $\mu$m long GKTFET shows much better $r_{out}$ and cutoff frequencies $f_{max}$ than GFETs due to the transmission gap engineered in pristine graphene using gate geometry. Perfect edges in a GKTFET would further reduce the leaked density of states in the transmission gap, leading to larger $r_{out}$ and $f_{max}$. The small contact resistances $R_S/R_D$ also have significant impacts to the cut off frequency$f_{T}$ as they compete with $r_{out}$.  Compared with GFET, the $f_T$ of the GKTFET is less sensitive to $R_S$ and $R_D$ due to a larger $r_{out}$.  
 
The parasitic capacitances $C_{gd}$ and gate resistance $R_G$ are critical device parameters and significantly impact $f_T$ and $f_{max}$. The $C_{gd}$ and $R_G$ are strongly dependent on the real design and geometry of the transistor. For instance, recent experiments used T-shape gate to reduce $R_G$ dramatically  to get high $f_{max}$ in GFETs \cite{guo2013record}. Simply including a sizeable side gate would create a large parasitic capacitance AC connected to the ground. To mitigate this, we will need to reference the third gate to the drain at a constant bias offset or include an inductor between the two to filter out the high-frequency AC signals. A more convenient choice would be to dope the two end regions chemically and work with a trapezoidal central gated region alone.

While our simulations were done for 1 $\mu$m, the ultimate advantage of the GKTFET for high-performance RF depends on its overall scalability, since the $C_{gs}$ and $R^{-1}_G$ are proportional to the channel length.
In the GKTFET, the gate width and length are related, as we used a $45^\circ$ tilted Junction. We show the $f_T$ and $f_{max}$ of GKTFETs and GFETs in Fig.\ref{fig:summary_of_fT_fmax}. Ideally, the $f_{T}$ and $f_{max}$ follow
$f_T \propto C_{gs}^{-1} \propto L^{-1}$ and $f_{max} \propto C^{-1}_{gs}R_G^{-0.5} \propto  L^{-0.5}$. The scaling of the gate length of the GKTFET with from 1 $\mu$m to 100 nm is expected to increase the $f_T$ and $f_{max}$ by 10 and 3.2 times respectively, as indicated by the dashed lines. In contrast, the $f_{max}$ of GFETs does not scale with channel length due to the low output resistance, as shown in Fig.\ref{fig:summary_of_fT_fmax} (f). To estimate the $f_T$ and $f_{max}$ of 100 nm GKTFETs, we made the following assumptions: the scaling down of the GKTFET does not change the electrostatics in the device (gate control is still dominant), the pseudogap can be effectively created
by GPNJs in a scaled GKTFET, and device parameters such as $C_{gs}$ and $C_{gd}$ scale properly with channel length {while maintaining the transition length (split length, d) across junctions in the range of 50-100 nm for better electron filtering resulting in transmission gap}. Detailed quantum simulations coupled with numerical 3D electrostatics are needed to test the performance of these devices at their scaling limits. {Trap charges, edge roughness, junction roughness, and contact resistance are important factors that will affect RF performances of realistic devices, therefore, those variabilities in real devices should be carefully calibrated\cite{xu2013variability}. In principle these non-idealities can be mitigated using hBN substrates, gated edges, graphite gates, and 1D metal edge contacts (contact resistance $\sim$150 $\Omega$-$\mu$m \cite{wang2013one}) which will be explored in future publications.}

\section{Conclusion}
To summarize, we propose a conceptual high-frequency RF device in Fig.\ref{fig:device}. This device operates by geometry engineering of a gate-tunable transport gap in pristine graphene, using the physics of Klein tunneling. In contrast to conventional GFETs which suffer from weak current saturation due to gaplessness, the engineering of a transmission gap allows the GKTFET to enjoy both high carrier lifetimes and current saturation. Our calculation of the GKTFET shows a significant improvement on $f_{max}$  and a slightly higher $f_T$ compared with GFETs. The device is expect to achieve an $f_T$ of 33 GHz and a comparable $f_{max}$ of 17 GHz in a device  with 1 $\mu$m gate length, and ramp up to $f_T $=350 GHz and $f_{max}$=53 GHz as we shrink the gate to 100 nm. Higher $f_{max}$ of 49 GHz for 1 $\mu$m channel and 158 GHz for 100 nm channel can be expected by reducing gate resistance  with the technique of T-Gate.
In addition, the cut-off frequencies of the GKTFET are seen to be much less sensitive to the contact resistance than GFETs, once again due to the significant increase in output resistance arising from current saturation.
%

\nocite{*}


\section*{Acknowledgments}
This work is supported in part by the Nanoelectronics Research Corporation (NERC), a wholly owned subsidiary of the Semiconductor Research Corporation (SRC), through the Institute for Nanoelectronics Discovery and Exploration (INDEX). The authors acknowledge computational resources on UVa HPC System Rivanna. We acknowledge Cory R. Dean from Columbia University, Kurt Gaskill from Naval Research Laboratory, Claire Berger from Georgia Institute of Technology and Philip Kim from Harvard University for helpful discussions.

\section*{Author contributions statement}
Y. T. analyzed the device DC and RF characteristics and wrote the manuscript.
M. E. performed the ray-tracing calculations for the current and wrote the manuscript. H. T. analyzed the device parasitic capacitors. K. H. and M. E. developed the ray-tracing code. 
N. B. and A. G. supervised the work. A. G. initialized the idea. All authors contributed to revising the manuscript.

\section*{Additional information}
Competing financial interests: The authors declare no competing financial interests.

\end{document}